\documentclass[review]{elsarticle}

\usepackage{graphicx}
\usepackage{subfig}
\usepackage{epstopdf}
\usepackage{amsmath}
\usepackage{dcolumn}
\usepackage{bm}
\usepackage[colorlinks=true, linkcolor=blue]{hyperref}

\journal{Journal of Hazardous Materials}

\begin{document}

\begin{frontmatter}

\title{Plasma filtering techniques for nuclear waste remediation}

\author[PPPL]{Renaud Gueroult\corref{RG}}
\ead{rgueroul@pppl.gov}

\author[SRNL]{David T. Hobbs}

\author[PPPL]{Nathaniel J. Fisch}

\address[PPPL]{Princeton Plasma Physics Laboratory, Princeton, New Jersey, 08540, USA}

\address[SRNL]{Savannah River National Laboratory, Aiken, South Carolina, 29808, USA}

\cortext[RG]{Corresponding author at: Princeton Plasma Physics Laboratory, P.O. Box 451, Princeton, New Jersey, 08540, USA. Tel.: +1 6092432493  Fax: +1 6092432662}

\begin{abstract}
Nuclear waste cleanup is challenged by the  handling of feed stocks that are both unknown and complex. 
Plasma filtering,  operating on dissociated elements, offers advantages over chemical methods in processing  such wastes. 
The costs incurred by plasma mass filtering for nuclear waste pretreatment, before ultimate disposal, are similar to those for chemical 
pretreatment.
However, significant savings might be achieved in minimizing the  waste mass. 
This advantage may be realized over a large range of chemical waste  compositions, thereby addressing the heterogeneity of legacy nuclear waste.  
\end{abstract}


\end{frontmatter}

\newpage

\section{Introduction}
\label{Sec:Introduction}

Stored radioactive waste proliferated with the development of nuclear weapons.  
Beginning with the Manhattan project, and throughout the cold war,  large quantities of radioactive waste were accumulated. 
Most of this waste originated as a byproduct of uranium and plutonium production at the Hanford and Savannah River sites, and from the enrichment plant at Oak Ridge~\cite{Crowley2002}. 
Before the 1970s, the composition of this waste was poorly documented, and significant quantities of liquid waste were released directly to the environment~\cite{Gephart2003}. 
Only the most highly radioactive fraction of the waste was piped to underground storage tanks.

At Savannah River, $36$ million gallons of high level waste are stored in $45$ underground tanks~\cite{Chew2014a}. 
Processing and immobilization of  high level waste in borosilicate glass started in $1996$. 
A salt waste processing facility is currently under construction, with first operations scheduled in $2018$. 
Completion of clean-up activities is scheduled only by $2033$~\cite{Chew2014b}. 

At Hanford, $54$ million gallons of waste were stored in $177$ underground tanks~\cite{Gephart2003,Certa2011}. 
The oldest, single shell, tanks were built between $1943$ and $1964$, with designed service lives of $10$ to $20$ years. 
Out of these $177$ tanks, $67$ have or are suspected to have leaked  up to $1$ million gallon into the environment~\cite{Gephart2003}, with first leaks confirmed in $1959$. 
Double shell carbon-steel tanks were built starting in $1968$ to provide better confinement.
Waste was then pumped from single shell to double shell tanks, yet, $2.8$ million gallons were still stored in single shell tanks in $2012$~\cite{Triplett2013}. 
Moreover,  leaks have also been discovered between shells of double shell tanks~\cite{DOE2012}. 
Construction of a facility to immobilize the high level waste using similar approaches to those used at Savannah River began in $2002$. 
However, due to various unresolved technical problems and work stoppages~\cite{Trimble2009}, the estimated cost to construct this treatment and immobilization facility has tripled from $4.3$ to $13.4$ billion dollars, and its scheduled completion date slipped by nearly a decade to 2019~\cite{Trimble2012}. 
Completion of clean-up activities is not expected before 2050~\cite{RiverProtection2013}. 
Clean up efforts for all the waste sites are projected to cost more than $280$ billion dollars~\cite{Friedman2014}. 

In essence,  clean-up  is a matter of  separating small volumes of high activity waste from much larger volumes of low activity waste. 
The separated high activity waste is then immobilized as glass for ultimate disposal in an underground repository. 
The low activity waste is immobilized in a less durable wasteform for onsite disposal. 

The presence of significant volumes of non-radioactive elements inside the high-activity waste stream is costly. 
First, vitrifying non-radioactive material incurs  the production cost of additional glass canisters, which  is a significant fraction of the total clean-up cost, since each canister costs on the order of a million dollars \cite{Bell1992,Swanson1993,NationalResearchCouncil1996,DeMuth1996,NationalResearchCouncil2001}. 
Moreover, the larger number of glass canisters requires a greater number of vitrification facilities,  increasing the capital cost. 
Second, the glass formulation has specific weight loading tolerances for different elements~\cite{NationalResearchCouncil2001,Kim2011}. 
For example, chromium, ruthenium, rhodium and palladium  in the glass can precipitate and eventually short circuit the glass melter electrodes. 
Furthermore, chromium, phosphorus oxide and sodium sulfate dissolve poorly in borosilicate glass, forming on occasions refractory crystalline phases that could compromise the durability of borosilicate glass wasteform. 

Thus,  the efficient separation of high-level radioactive elements from the low-level waste can lower significantly the cost of the clean-up~\cite{Bell1992,NationalResearchCouncil1996}.

It is the objective here to examine the practicality, or economic feasibility, specifically of plasma mass filtration techniques for nuclear waste clean-up.  
In doing so, it is our further objective to identify those tasks that might best be accomplished by plasma-based techniques when used together with other techniques. 

The utility of plasma-based techniques depends on the nature of the nuclear waste, which is often highly heterogeneous.
It is also often the case that elements of very different atomic weights require separation.
We will show that it is on these types of wastes that plasma techniques tend to be economically competitive.

This  paper is organized as follows:
 In Section~\ref{Sec:Challenges}, we examine the main challenges faced by waste tank clean-up operations.  
 We  take the Hanford waste as an example, which illustrates certain  limitations to chemical techniques. 
 In Section~\ref{Sec:Plasma_filtering}, we review the essential characteristics of plasma mass filtering techniques. 
 In Section~\ref{Sec:Economics}, we compare the projected costs of plasma techniques to costs for  chemical techniques for the particular application of sludge pretreatment. 
 In Section~\ref{Sec:Summary}, we summarize the main results.

\section{Tank clean-up challenges}
\label{Sec:Challenges}

Although conceptually simple, separating non-radioactive material from radioactive elements can prove to be extremely challenging.
In the case of legacy waste analyzed here, the challenge arises from the heterogeneity of the input stream, both in terms of physical and chemical forms. 
Waste stored in tanks is in one of three forms~\cite{NationalResearchCouncil1996,Nazarro2003}. 
Due to the high pH, the bulk of the metals precipitate as insoluble metal oxides/hydroxides that gravity settles to form a thick layer referred to as \emph{sludge}. 
Typical metals include Al, Bi, Cr, Fe, Mn, Si and U. The liquid fraction of the waste, referred to as \emph{supernate}, contains water-soluble components, principally the sodium salts of oxyanions including hydroxyde, nitrate, nitrite, aluminate, sulfate and carbonate. 
Historically, the supernate has been evaporated to minimize the volume. 
Cooling of the hot, concentrated supernate produced crystalline salts, which accumulated in a layer referred to as \emph{saltcake}.

Typical waste pretreatment operations can be summarized as follows~\cite{Carreon2002}. 
The sludge is recovered and goes through a series of caustic leaching, oxydative leaching and washing steps to remove non-radioactive elements, in particular Na, Al and Cr~\cite{Fiskum2009}. 
Saltcake is dissolved in water and combined with supernates and liquids from sludge leaching and washing. 
Undissolved solids are removed and the clarified liquids are treated to remove certain radionuclides such as $^{137}$Cs, $^{99}$Tc and $^{90}$Sr. The leached and washed sludge, together with the elements removed from the dissolved saltcake and supernate, are combined and vitrified. After the removal of the radionuclides, the decontaminated supernate is immobilized in either cementitious (Savannah River Site) or glass (Hanford site) wasteform.

\begin{figure}[htpb]
\begin{center}
\includegraphics[width=14cm]{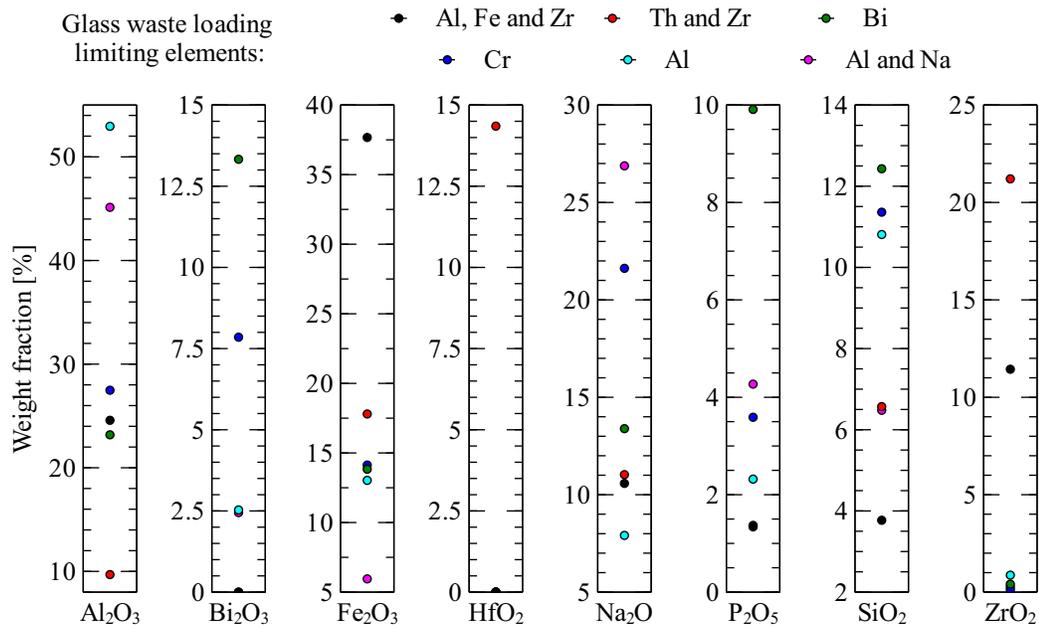}
\caption{Oxide waste mass breakdown across the six different sub-groups of Hanford wastes presenting challenges in terms of canister waste loading and glass formulation, from~\cite[Table 2.2]{Kim2011}. Only oxides with mass fraction over $10\%$ are plotted here. }
\label{Fig:composition} 
\end{center}
\end{figure}

In the case of Hanford,  the large waste compositional variations between tanks complicates the separation process~\cite{Snow2007,Kim2011}, as illustrated in Fig.~\ref{Fig:composition}. 
The high level waste at Hanford can be divided into six sub-groups based on their chemistry and glass formulation limiting factors~\cite{Vienna2014}: high alumina wastes, high iron wastes, high iron, chromium, nickel and manganese wastes, high chromium and sulfur wastes, high phosphorus and calcium wastes, and high alkali wastes. 
Removal of non radioactive elements by means of chemical techniques is then  challenging,  since the elements to be removed vary widely from batch to batch, and are usually a combination of elements with various chemical properties. 
As a result, chemical separation (\emph{e.~g.} aluminium in the chemical form of boehmite~\cite{Smith2011}, and chromium present as Cr(III) compounds~\cite{Sylvester2001,Lumetta2008}) has proven to be particularly difficult~\cite{Fiskum2009}.

To avoid these difficulties, new glass formulation methods may allow higher aluminum and chromium fractions as well as higher waste loadings~\cite{Kruger2010,Kruger2011}. 
These methods may stem the increase of canisters resulting from larger volumes, but the  glass formulation is difficult and still uncertain. 
The capability to accommodate typically encountered waste composition variations is yet to be demonstrated. 
In addition, higher aluminum content will most likely have detrimental side effects, notably on achievable processing rates~\cite{Pierce2012}, and may call for different melter technology solutions~\cite{Smith2014}. 


Yet another approach is to reduce cost through pretreatment of the waste~\cite{Aloise2009}.
Here non-chemical separation techniques are attractive since they are in principle indifferent to waste heterogeneity. 
One example of such a non-chemical technique is plasma mass filtering.

\section{Plasma mass filtering}   
\label{Sec:Plasma_filtering}

The potential of plasma medium to separate elements based on their mass has long been recognized~\cite{Lehnert1971}. 
An example of such a device is the plasma centrifuge~\cite{Krishnan1981}, which operates in a similar fashion to conventional gaseous or liquid centrifuges, but offers higher separation factors due to its ability to operate at much larger rotation speeds. 
Higher rotation speeds are in this case made possible by the absence of moving parts, with rotation produced in this device by means of the combined effects of electric and magnetic fields~\cite{Lehnert1971}. However, the main thrust for this research effort was originally isotope separation~\cite{Grossman1991,Rax2007}. 
As a result, most of the work was directed towards low mass differences and, consequently, low throughput. Only recently has plasma mass filtering been considered for nuclear waste remediation~\cite{Freeman2003} and for nuclear spent fuel reprocessing~\cite{Gueroult2014a}. 
The use of plasmas for these new applications was made possible by the development of various new plasma filter concepts~\cite{Ohkawa2002,Fetterman2011,Gueroult2012a,Gueroult2014} which offer high-throughput processing granted sufficiently large mass differences between species to be separated~\cite{Fetterman2011b}. 

In these devices, material can be fed in the machine in different forms. 
Possible candidates include powder injection or laser evaporation. 
Although the choice of a particular feeding technique has not been made yet, and will most likely depend on the specifics of the targeted process, general constraints can be obtained for this particular process. For example, in the case of powder injection, micron-size particles are likely to be required for the envisioned plasma operating conditions~\cite{Tanaka2007}. 
Similarly, the desired throughput will dictate the required laser power. 

\begin{figure}[htpb]
\begin{center}
\subfloat[~Archimedes Filter~\cite{Freeman2003}]{\includegraphics[]{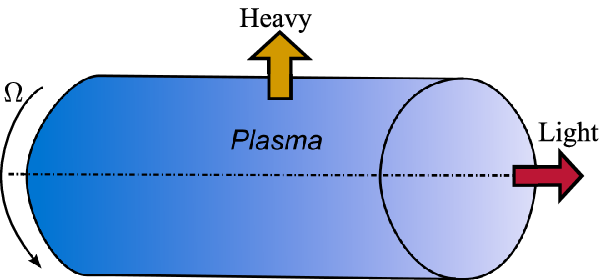}\label{Fig:Archimedes}}$\quad$\subfloat[~Double Well Filter~\cite{Gueroult2014}]{\includegraphics[]{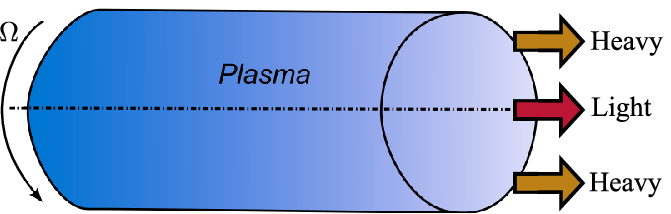}\label{Fig:DoubleWell}}\hspace{0.1cm}
\subfloat[~MCMF Filter~\cite{Fetterman2011}]{\includegraphics[]{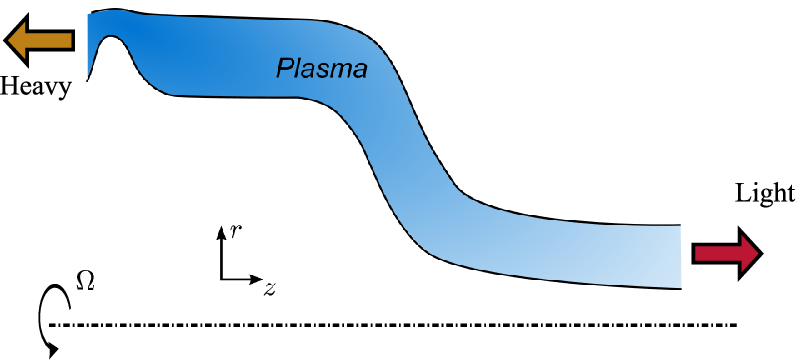}\label{Fig:MCMF}}
\caption{Different high throughput plasma filter concepts: axial/radial separation in the Archimedes filter \protect\subref{Fig:Archimedes}, radial layering in the Double Well filter \protect\subref{Fig:DoubleWell} and axial/axial separation in the MCMF filter \protect\subref{Fig:MCMF}. These three filter concepts feature axisymmetric rotating plasmas. }
\label{Fig:plasma_filters} 
\end{center}
\end{figure}

Once ionized, charged particles respond to both electromagnetic and centrifugal fields. In plasma filters devices, these fields are generally designed such that there exists a mass threshold $m_c$ for particle confinement. Elements heavier than the mass threshold $m_c$ are then directed one way, while elements lighter than this mass threshold are directed in another way. Fig.~\ref{Fig:plasma_filters} illustrates the differential confinement properties of \emph{light} and \emph{heavy} elements for the three main filter concepts. In is worth noting here that variations on these concepts exist, such as the use of RF electric fields in place of DC electric fields controlling the plasma rotation. This could in principle allow isolating a particular mass from the bulk~\cite{Ohkawa2002}, rather than discriminating elements based on a threshold mass.

The two separated streams can then be recovered individually. 
Depending on the selected filter concept, charged particles could be either deposited and neutralized on a surface, or neutralized in volume by locally tuning the plasma parameters. 

\section{Economic feasibility of plasma filtering techniques for sludge pretreatment}   
\label{Sec:Economics}

\begin{figure}[htpb]
\begin{center}
\includegraphics[]{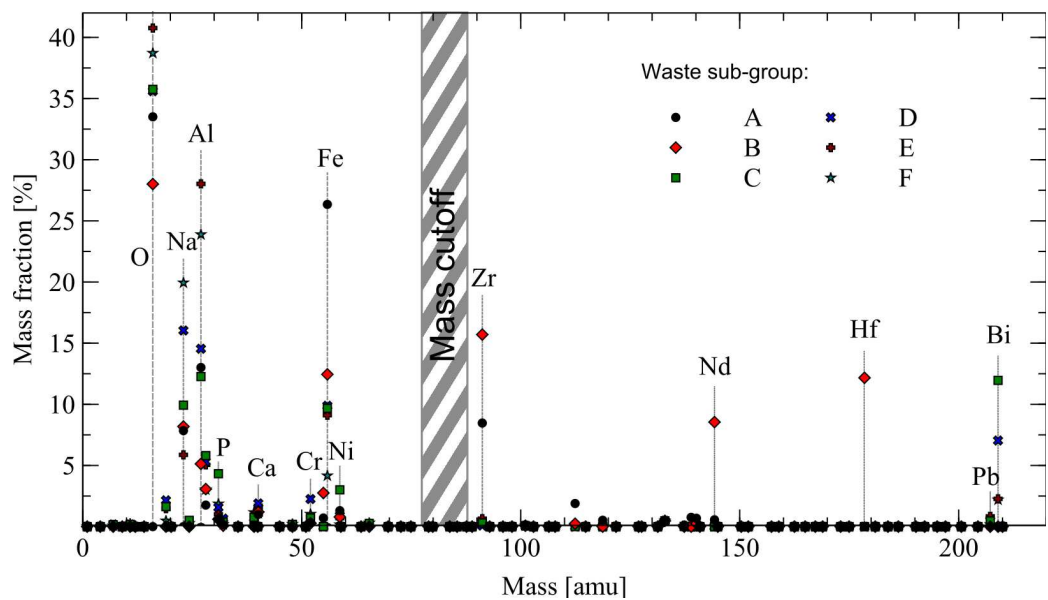}
\caption{Waste weight composition by elements, from~\cite[Table 2.2]{Kim2011}. The plasma mass filter cutoff mass is indicated by the shaded box. The different waste compositions correspond to the different sub-groups summarized in Tab.~\ref{Tab:tab0}. }
\label{Fig:CompoByMass} 
\end{center}
\end{figure}

\begin{figure}[htpb]
\begin{center}
\subfloat[Chemical flowchart]{\includegraphics[]{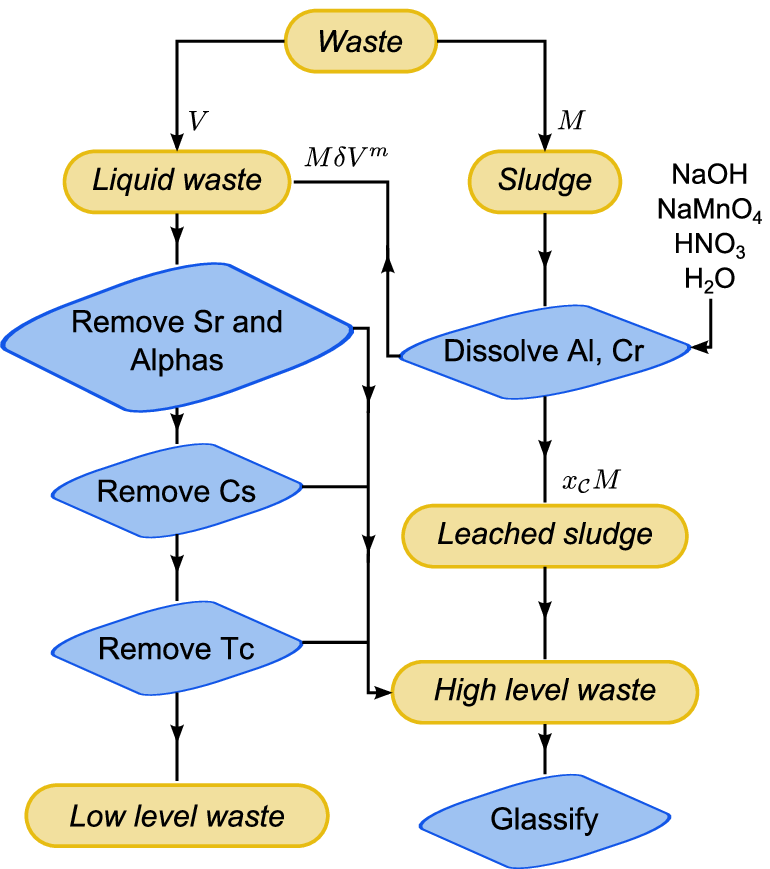}\label{Fig:Chemical}}\subfloat[Plasma flowchart]{\includegraphics[]{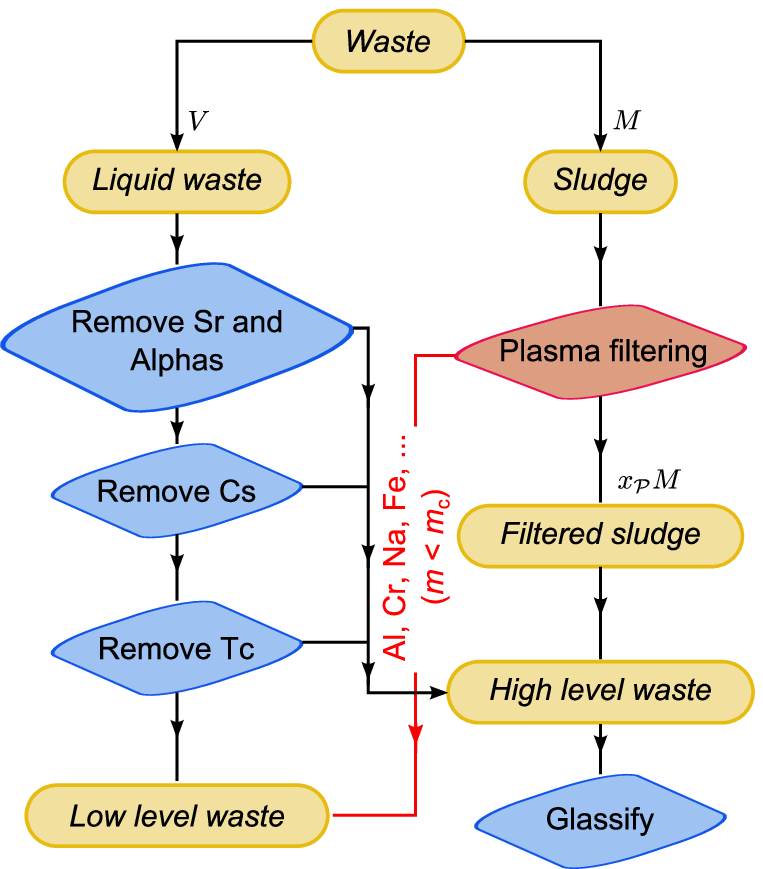}\label{Fig:Plasma}}
\caption{Chemical~\protect\subref{Fig:Chemical} and plasma~\protect\subref{Fig:Plasma} flowcharts for waste processing. Plasma approach could in principle suppress the additional liquid waste produced by sludge washing and leaching, as well as minimize the final volume of pretreated sludge to be vitrified. }
\label{Fig:Flowchart} 
\end{center}
\end{figure}

Although different insertion points can be envisioned, the ability to separate a \emph{light population} from a \emph{heavy population} makes plasma filters attractive for sludge pretreatment. For example, for the typical sludge composition introduced in Sec.~\ref{Sec:Challenges} and plotted in Fig.~\ref{Fig:CompoByMass}, one can imagine tuning the plasma filter in such a way that Al, Cr, Fe, O, Na and Si are below the cutoff mass, while Sr, Tc, Cs, Bi, Th and U are heavier than the cutoff mass. As indicated in Fig.~\ref{Fig:Plasma}, the low volume, heavy stream, could then be processed as high activity waste and vitrified together with radionuclides recovered from the liquid waste, while the larger volume, light stream will be processed as low activity waste. The total waste mass below this cutoff mass, as summarized in Tab.~\ref{Tab:tab0}, gives an upper limit for the plasma treatment efficiency of $60\%$ to $95\%$. 

\begin{table}
\begin{center}
\begin{tabular}{c | c | c}
Waste sub-group & Limiting glass factor & Mass fraction under $90$~amu [\%]\\
\hline
A & Al, Fe and Zr & $86.5$\\ 
B & Th and Zr & $ $$62.5$\\
C & Bi & $86.7$\\
D & Cr & $91.8$\\
E & Al & $95.5$\\
F & Al and Na & $96.8$\\
\end{tabular}
\caption{Waste subgroups by limiting factor for vitrification, from~\cite{Kim2011}, and computed mass fraction under the mass filter threshold.}
\label{Tab:tab0}
\end{center}
\end{table}

More generally, since the common pattern is to separate heavy radioactive elements from lighter non-radioactive elements, the plasma filters could be tuned to respond best to a given waste composition. It is worth noting here that, as opposed to chemical techniques, such a tuning could in principle be done on the fly as it would essentially consist in setting the rotation speed accordingly. Beyond the rotation speed control achieved through the transverse electric field, other plasma parameters, such as electron and ion temperatures and background neutral pressure, can be modified to optimize the separation efficiency~\cite{Gueroult2012a}.  

\subsection{Cost of chemical sludge disposal}

Looking at the chemical processing flowchart depicted in Fig.~\ref{Fig:Chemical}, the cost of sludge disposal per unit mass ${{\mathcal{C}}_{\mathcal{C}}}^m$ can be broken down to the sum of the individual cost of three subprocesses,
\begin{equation}
{{\mathcal{C}}_{\mathcal{C}}}^m = {\mathcal{C}_{s}}^{m} + x_{\mathcal{C}}{\mathcal{C}_{v}}^{m} +  \delta V^m{\mathcal{C}_{l}}^{V}, 
\end{equation}
where ${\mathcal{C}_{s}}^{m}$ is the cost of sludge washing and leaching per unit mass, ${\mathcal{C}_{v}}^{m}$ is the vitrifying cost per unit mass of waste load, $x_{\mathcal{C}}$ is the mass of solid waste after washing and leaching one kg of sludge and $\delta V {\mathcal{C}_{l}}^{V}$ is the cost of additional liquid waste processing (${\mathcal{C}_{l}}^{V}$ is the liquid waste processing cost per unit volume, and $\delta V^m$ is the volume of liquid waste produced by washing and leaching of a kilogram of sludge). To be exhaustive, one would also have to account for the cost associated with the disposal of solid waste generated during the sludge pretreatment, as well as during the removal of radionuclides in the additional liquid waste. This additional solid waste will be combined with the sludge for vitrification, and will consequently result in an incremental increase of $x_{\mathcal{C}}$. However, since only limited amount are expected to derive directly from sludge washing and leaching, additional solid wastes are neglected in this study.   

Pretreatment costs estimates can be inferred from previous studies, and are summarized in Tab.~\ref{Tab:tab1}. Corrected for inflation, pretreatment costs are respectively \$$24$ and \$$43$ per kg of liquid and sludge waste~\cite{McGinnis1999}. Assuming that liquid waste is essentially made of sodium hydroxide ($\rho_{\textrm{NaOH}}=2.13$~g.cm$^{-3}$), this gives ${\mathcal{C}_{l}}^{V}\sim\$50~$ per liter of liquid waste and ${\mathcal{C}_{s}}^{m}\sim\$45$ per kg of sludge. Moving to vitrification, the cost per unit mass can be estimated from the published incremental cost of producing one more, or one fewer, canister~\cite{Perez2001}. Corrected again for today's dollar, this incremental cost is estimated between \$$0.8~$M and \$$1.5~$M, with about half of it resulting from storage. Using a standard $2~$ft$\times10~$ft glass canister~\cite[Appendix E]{NationalResearchCouncil1996},  densities of $2.6$ for the glass~\cite[p.~9]{DeMuth1996} and $5$ for the waste ($\rho_{Al}\sim2.7$~g.cm$^{-3}$, $\rho_{Fe}\sim7.9$~g.cm$^{-3}$), and a $25\%$ waste weight loading gives a vitrification cost per unit mass ${\mathcal{C}_{v}}^{m}$ between $\$1200$ and \$$2300$. Finally, sludge washing is responsible for an additional $1.3$ liter of salt waste per liter of sludge processed~\cite[p.~37]{Chew2014}. Using here aluminum density as a baseline for sludge, this gives $\delta V^m \sim 0.5$~liter per kg of sludge.

\begin{table}
\begin{center}
\begin{tabular}{c | c | c}
& Subprocess & Cost estimate\\
\hline
${\mathcal{C}_{s}}^{m}$ & Sludge washing and leaching & \$$45$ per kg of sludge\\ 
${\mathcal{C}_{v}}^{m}$ & Vitrification\protect\footnotemark & $ $\$$1200-2300$ per kg of waste load\\
$\delta V^m$ & Additional liquid waste produced & $0.5~$L per kg of sludge\\
${\mathcal{C}_{l}}^{V}$ & Liquid waste treatment & \$$50$ per liter of liquid waste\\
\end{tabular}
\caption{Breakdown of chemical processing costs.}
\label{Tab:tab1}
\end{center}
\end{table}
\footnotetext{This includes the canisters storage cost. }

Overall, the total chemical sludge disposal cost stems essentially from vitrification and glass canister storage costs, while pretreatment costs are negligible in comparison.

\subsection{Cost of plasma assisted sludge disposal}

Looking back at the chemical and plasma flowcharts in Fig~\ref{Fig:Flowchart}, one sees that a plasma approach would eliminate the secondary liquid waste stream ($\delta V^m\sim0$). The cost of plasma sludge disposal ${{\mathcal{C}}_{\mathcal{P}}}^m$ per unit mass of sludge is thus the sum of only two subprocesses, 
\begin{equation}
{{\mathcal{C}}_{\mathcal{P}}}^m = x_{\mathcal{P}}{\mathcal{C}_{v}}^{m} + {\mathcal{C}_{pf}}^{m} . 
\end{equation}
The first one is the vitrifying cost, where ${\mathcal{C}_{v}}^{m}$ is the vitrification cost per unit mass of waste load, and $x_{\mathcal{C}}$ is the mass of solid waste after plasma filtering. ${\mathcal{C}_{v}}^{m}$ is identical to the one obtained for chemical separation and listed in Tab.~\ref{Tab:tab1}, and only $x_{\mathcal{P}}$ differs. The second one, ${\mathcal{C}_{pf}}^{m}$, is the cost of plasma filtering per unit mass. This plasma filtering cost, can itself be broken down into different processes.

\paragraph{Evaporation}

First, the waste needs to be fed into the machine. As discussed in Sec.~\ref{Sec:Plasma_filtering}, one option consists in laser evaporation. Assuming that the latent heat of vaporization $\mathcal{L}_v$ is dominant over both the latent heat of fusion and the enthalpy change due to the temperature increase, an firth order estimate for the evaporation cost ${\mathcal{C}_e}^m$ is $\mathcal{L}_v/\chi$, where $\chi$ is the laser absorptivity. For an aluminum rich waste, $\chi\sim0.2$~\cite{Mazhukin2007} and ${\mathcal{L}_v}^{\textrm{Al}}\sim10~$MJ/kg, ${\mathcal{L}_v}^{\textrm{Al$_2$O$_3$}}\sim4.8~$MJ/kg~\cite[p. 115]{Samsonov1973}, gives ${\mathcal{C}_e}^m\sim20-50~$MJ/kg.

\paragraph{Plasma production}

Once the waste turned into a gas, the next step consists in ionizing this gas. Using once more aluminum as a baseline, this requires $21.4~$MJ/kg. This figure is again an ideal value. In practice, one has to account for all energy dissipation channels. First, part of the electron energy will be dissipated through excitation of neutrals and ions. A measure of the deviation from the ideal case is the efficiency $\eta$, defined as the ratio of the energy required for one electron-ion pair creation over the atom ionization energy $\varepsilon_i$. For helicon discharges envisioned for this application, $\eta$ is about $0.4$ for an electron temperature $T_e\sim4~$eV in pure Argon ($\varepsilon_i=15.75~$eV)~\cite[p. 81]{Lieberman1994}. However, the efficiency $\eta$ is expected to be reduced for the more complex compositions typically envisioned here. It is worth noting here that since excitation losses scale with the square of the plasma density, $\eta$ can in principle be maintained to acceptable levels by limiting the plasma density. 

In addition to electron losses, other energy dissipation channels might have to be considered depending on the plasma parameters. These includes the energy transfer to ions in the form of rotational kinetic energy and temperature. Quantitatively, a mass of $1~$kg rotating at $3~$km.s$^{-1}$ has a kinetic energy of $4.5~$MJ, and increasing the temperature of $1$~kg of aluminum gas by $1~$eV requires $3.3~$MJ. These losses are therefore small in most cases compared to electron losses. Assuming a highly degraded $\eta\sim0.02$, the cost of plasma formation and maintenance is of the order of $1~$GJ/kg. 

\paragraph{Total cost}
Summing up the costs of these two sub-processes, and assuming a low laser electric efficiency of $0.1$, gives a plasma filtering energy cost of the order of $1.5~$GJ/kg. Using a typical electricity cost of \$$0.1$ per kilowatt-hour (kWh), \emph{i.~e.} $36~$MJ/\$, this puts the cost of plasma pretreatment ${\mathcal{C}_{pf}}^{m} \sim \$40~$per kg of sludge. Interestingly, this cost is on par with ${\mathcal{C}_{s}}^{m}$, the cost of sludge chemical washing and leaching. 

The absence of secondary liquid waste stream would represent a saving of about \$$25$ per kg of sludge. However, the largest opportunity to reduce costs lies in waste mass minimization. As a matter of fact, because of the significant costs associated with vitrification, a higher mass minimization $x_{\mathcal{P}}<x_{\mathcal{C}}$ could offer large savings. 

\subsection{Waste mass minimization: possible savings}

Studies indicate high variations of chemical washing/leaching efficiency across the various tank farms~\cite{Harrington2011}, with aluminum removal values ranging from $20\%$ to $99\%$. Similar variations can be found for other elements of critical importance such as chromium and iron~\cite{Rapko2002}, and studies targeted to one particular waste type showed comparable results~\cite{Geeting2005}. As a consequence, non-radioactive elements are still largely responsible of the large mass of the processed sludge. Improvements in the chemical processes are challenging since further washing/leaching puts additional constraints on the liquid waste stream, and because a given process might increase the performances with respect to one specific element while degrading the performances with respect to another. On the other hand, a plasma mass filter could in principle offer on average higher efficiencies since all elements lighter than the mass threshold would be removed. 

\begin{figure}[htpb]
\begin{center}
\includegraphics[]{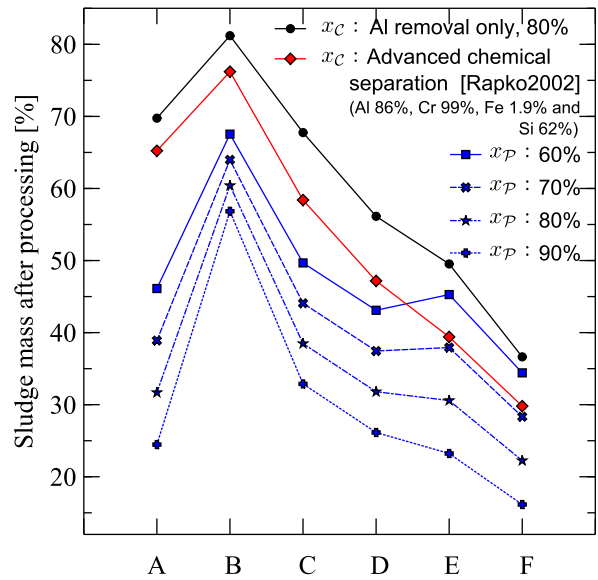}
\caption{Sludge mass decrease per waste sub-group (A to F, see Tab.~\ref{Tab:tab0}) as obtained by different pretreatment options and different plasma filtering efficiencies. $x_{\mathcal{P}}$ and $x_{\mathcal{C}}$ denote the mass fraction of sludge after respectively plasma and chemical processing. }
\label{Fig:Separation} 
\end{center}
\end{figure}

Using the the typical waste subgroups compositions depicted in Fig.~\ref{Fig:CompoByMass}, one can produce waste mass minimization estimates for various separation schemes. Three different separation processes are studied here. The first one consists simply in the removal of $80\%$ of the aluminum oxide contained in the sludge. The second one corresponds to an optimized chemical separation process~\cite{Rapko2002}, with the removal of $86\%$ of Al$_2$O$_3$, $99\%$ of Cr$_2$O$_3$, $1.9\%$ of Fe$_2$O$_3$ and $62\%$ of SiO$_2$. The third one corresponds to a plasma filtering with a threshold mass $m_c=90~$amu, and a given uniform separation efficiency for all oxides for which the non oxygen element is lighter than $m_c$. In all three cases, sodium is assumed to be totally removed during the process. The corresponding results are plotted in Fig.~\ref{Fig:Separation}. As expected, there are large variations depending on waste composition, and that both for plasma and chemical techniques. Performance is the worst for the high thorium and zirconium wastes (sub-group B), which makes sense since these elements are not recovered by either of these processes. Comparing chemical and plasma techniques, it appears that a moderate $60\%$ plasma filter separation efficiency leads  to comparable or better results than the advanced chemical process. The advantage of plasma filtering decreases for high aluminum content wastes (sub-groups E and F), since in this case the higher aluminum separation efficiency offered by chemical techniques addresses better the problem. However, a $70\%$ plasma filtering separation efficiency is sufficient to make the plasma approach more efficient than the chemical process modeled across all waste sub-groups. 

\begin{figure}[htpb]
\begin{center}
\includegraphics[]{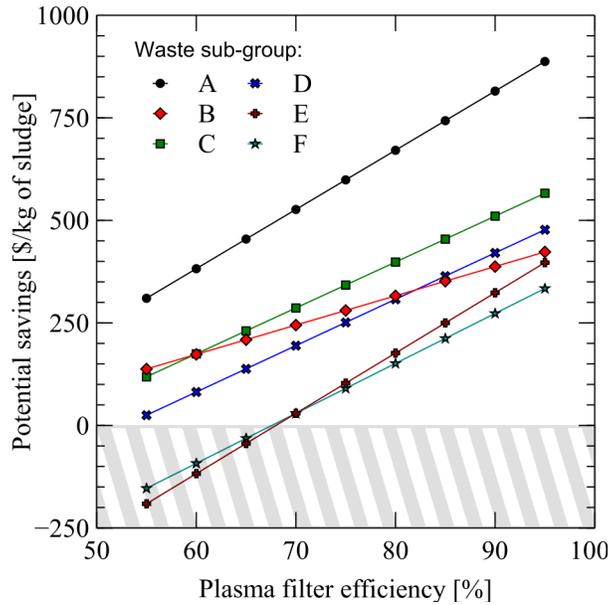}
\caption{Projected savings by kg of sludge offered by plasma processing due to waste mass minimization, $(x_{\mathcal{C}}-x_{\mathcal{P}}){\mathcal{C}_{v}}^{m}$. A typical advanced chemical process~\cite{Rapko2002} is used as the baseline for this comparison (removal of $86\%$ Al, $99\%$ Cr, $1.9\%$ Fe and $62\%$ Si). ${\mathcal{C}_{v}}^{m} = 2000$\$/kg.  }
\label{Fig:Savings} 
\end{center}
\end{figure}

Waste mass minimization can be directly translated into cost savings using the typical vitrification cost per kg of processed sludge listed in Tab.~\ref{Tab:tab1}. The cost difference per kg of sludge due to waste mass minimization is $(x_{\mathcal{C}}-x_{\mathcal{P}}){\mathcal{C}_{v}}^{m}$. The corresponding data is plotted in Fig.~\ref{Fig:Savings}, and suggests that savings of \$$80$ and higher per kg of sludge are possible for two third of the waste types and a $60\%$ plasma mass filtering efficiency. Increasing the efficiency to $80\%$ yields savings of \$$150$ to \$$650$ per kg depending on the waste type. These values are substantial, representing $7.5$ to $32.5\%$ of the total sludge processing cost.     

It is worth noting here that a filtering efficiency of $70\%$ appears to be well within reach of the proposed plasma filter concepts~\cite{Gueroult2014a}. However, if required, higher values could be achieved by staging the filter, so that particles go through multiple separation steps. Three passes, each at an efficiency of $70\%$, would offer a $97\%$ separation. This could ideally be achieved at negligible cost by maintaining elements ionized throughout the full cycle. Even for the worst case scenario where particles have to be re-ionized, the cost per kg would not exceed two or three ${\mathcal{C}_{pf}}^{m}$. This is about \$$120$ per kg, which would be lower than the savings achieved as a result of an improved waste mass minimization.

To summarize, the cost of plasma filtering appears to be about the same as the costs associated with  chemical washing and leaching for sludge pretreatment. 
However, because no additional liquid waste produced, and because non-radioactive elements critical to the vitrification process are removed easily, plasma filtering may offer significant savings when considering the entire sludge pretreatment and vitrification process. .
This is especially true for highly heterogeneous waste, such as legacy waste.

\section{Summary}   
\label{Sec:Summary}

The cost of plasma mass filtering was analyzed  within the  generally accepted framework in which the ultimate disposal of the nuclear waste consists of immobilization of the radioactive components in glass for permanent storage in a geological repository. 
However, the  cost of vitrification rises with the waste mass, sometimes to a prohibitive level. 
This is especially true for legacy waste, that is to say nuclear waste produced as a byproduct of nuclear weapons development during the cold war era, which is typically made of large volumes of non-radioactive material mixed with much smaller volumes of highly radioactive elements. 
Disposal of this kind of waste hence requires efficient ways of separating radioactive and non-radioactive components, in order to minimize the mass of the high-activity fraction.

The large number of chemical elements present in the waste challenges chemical separation approaches.
Significant variations both in physical and chemical forms further challenges the chemical techniques. 
In contrast, plasma mass filtering techniques appear promising because of their ability to discriminate elements irrespective of their chemical composition. 
For  applications to nuclear waste, plasma filtering can exploit the large mass gap existing between most of the non radioactive elements and the smaller fraction of radioactive elements.

We estimated that, for Hanford wastes,  the processing costs for chemical and plasma filtering techniques are comparable. 
However,  plasma processing could, in principle, provide for significantly higher reduction in the mass of the waste sent to the high-activity vitrification melter compared to chemical techniques. 
The reduction in high-activity waste would yield significant savings. 
For example, a plasma filter offering $70\%$ separation efficiency for the non-radioactive elements would decrease the overall cost by \$$30$ to \$$530$ per kg of sludge, which represents a $1.5$ to $26.5\%$ savings. 
In addition, as opposed to chemical washing and leaching, plasma sludge processing does not produce additional liquid waste that would require additional treatment.   

Thus, our preliminary evaluation suggests the economic feasibility of plasma filtering for sludge pretreatment. 
It remains, however, to refine our cost comparison by including capital, operation and maintenance costs. 
It remains also to analyze how waste mass minimization offered by plasma filtering might be advantageously combined with advanced glass formulations for increased waste loadings and reduced number of glass canisters.

\section*{Acknowledgements}
This work was supported under DoE Contract Number DE-AC02-09CH11466.

\section*{References}

\begin{thebibliography}{51}
\expandafter\ifx\csname natexlab\endcsname\relax\def\natexlab#1{#1}\fi
\providecommand{\url}[1]{\texttt{#1}}
\providecommand{\href}[2]{#2}
\providecommand{\path}[1]{#1}
\providecommand{\DOIprefix}{doi:}
\providecommand{\ArXivprefix}{arXiv:}
\providecommand{\URLprefix}{URL: }
\providecommand{\Pubmedprefix}{pmid:}
\providecommand{\doi}[1]{\href{http://dx.doi.org/#1}{\path{#1}}}
\providecommand{\Pubmed}[1]{\href{pmid:#1}{\path{#1}}}
\providecommand{\bibinfo}[2]{#2}
\ifx\xfnm\relax \def\xfnm[#1]{\unskip,\space#1}\fi
\bibitem[{Crowley and Ahearne(2002)}]{Crowley2002}
\bibinfo{author}{K.~Crowley}, \bibinfo{author}{J.~F. Ahearne},
  \bibinfo{journal}{American Scientist} \bibinfo{volume}{90}
  (\bibinfo{year}{2002}) \bibinfo{pages}{514}.
  \DOIprefix\doi{10.1511/2002.6.514}.
\bibitem[{Gephart(2003)}]{Gephart2003}
\bibinfo{author}{R.~E. Gephart}, \bibinfo{title}{A Short History of Hanford
  Waste Generation, Storage, and Release}, \bibinfo{type}{Technical Report}
  \bibinfo{number}{PNNL-13605, Rev. 4}, Pacific Northwest National Laboratory,
  \bibinfo{year}{2003}.
\bibitem[{Chew(2014)}]{Chew2014a}
\bibinfo{author}{D.~P. Chew}, \bibinfo{title}{Savannah River Site - Waste Tanks
  Levels}, \bibinfo{type}{Technical Report}
  \bibinfo{number}{SRR-LWP-2010-00001, rev. 48}, Savannah River Site,
  \bibinfo{year}{2014}.
\bibitem[{Chew and Hamm(2014)}]{Chew2014b}
\bibinfo{author}{D.~P. Chew}, \bibinfo{author}{B.~A. Hamm},
  \bibinfo{title}{Liquid Waste System Plan Revision 19},
  \bibinfo{type}{Technical Report} \bibinfo{number}{SRR-LWP-2009-00001,
  rev.19}, Savannah River Site, \bibinfo{year}{2014}.
\bibitem[{Certa et~al.(2011)Certa, Empey, and Wells}]{Certa2011}
\bibinfo{author}{P.~J. Certa}, \bibinfo{author}{P.~A. Empey},
  \bibinfo{author}{M.~N. Wells}, \bibinfo{title}{River Protection Project
  System Plan}, \bibinfo{type}{Technical Report} \bibinfo{number}{ORP-11242,
  Rev. 6}, Office of River Protection, \bibinfo{year}{2011}.
\bibitem[{Triplett et~al.(2013)Triplett, Watson, and Wellman}]{Triplett2013}
\bibinfo{author}{M.~B. Triplett}, \bibinfo{author}{D.~J. Watson},
  \bibinfo{author}{D.~M. Wellman}, \bibinfo{title}{Risks from Past, Current,
  and Potential Hanford Single Shell Tank Leaks}, \bibinfo{type}{Technical
  Report} \bibinfo{number}{PNNL-22483}, Pacific Northwest National Laboratory,
  \bibinfo{year}{2013}.
\bibitem[{DOE(2012)}]{DOE2012}
in: \bibinfo{booktitle}{U. S. Department of Energy press release}. \URLprefix
  \url{http://energy.gov/em/articles/hanford-determines-double-shell-tank-leaked-waste-inner-tank}.
\bibitem[{Trimble(2009)}]{Trimble2009}
\bibinfo{author}{D.~C. Trimble}, \bibinfo{title}{Nuclear and Worker Safety:
  Limited Information Exists on Costs and Reasons for Work Stoppages at DOE's
  Hanford Site}, \bibinfo{type}{Technical Report} \bibinfo{number}{GA0-09-451},
  United States Governement Accountability Office, \bibinfo{year}{2009}.
  \URLprefix \url{http://www.gao.gov/products/GAO-09-451}.
\bibitem[{Trimble(2012)}]{Trimble2012}
\bibinfo{author}{D.~C. Trimble}, \bibinfo{title}{Hanford Waste Treatment Plant:
  DOE Needs to Take Action to Resolve Technical and Management Challenges},
  \bibinfo{type}{Technical Report} \bibinfo{number}{GAO-13-38}, United States
  Governement Accountability Office, \bibinfo{year}{2012}. \URLprefix
  \url{http://www.gao.gov/products/GAO-13-38}.
\bibitem[{{Office of River Protection}(2013)}]{RiverProtection2013}
\bibinfo{author}{{Office of River Protection}}, \bibinfo{title}{Hanford Cleanup
  Completion Framework}, \bibinfo{type}{Technical Report}
  \bibinfo{number}{DOE/RL-2009-10, Rev. 1}, U.S. Department of Energy,
  \bibinfo{year}{2013}.
\bibitem[{Friedman(2014)}]{Friedman2014}
\bibinfo{author}{G.~H. Friedman}, \bibinfo{title}{Special Report - Management
  Challenges at the Department of Energy ? Fiscal Year 2015},
  \bibinfo{type}{Technical Report} \bibinfo{number}{DOE/IG-0924}, U.S.
  Department of Energy, Office of Inspector General, Office of Audits and
  Inspections, \bibinfo{year}{2014}.
\bibitem[{Bell and Bell(1992)}]{Bell1992}
\bibinfo{author}{J.~T. Bell}, \bibinfo{author}{L.~H. Bell}, in:
  \bibinfo{editor}{W.~W. Schulz}, \bibinfo{editor}{E.~P. Horwitz} (Eds.),
  \bibinfo{booktitle}{Proceedings of the American Chemical Society Symposium on
  Chemical Pretreatment of Nuclear Waste for Disposal},
  \bibinfo{publisher}{Springer}, \bibinfo{year}{1992}, pp.
  \bibinfo{pages}{1--16}.
\bibitem[{Swanson(1993)}]{Swanson1993}
\bibinfo{author}{J.~L. Swanson}, \bibinfo{title}{Clean option: An alternative
  strategy for Hanford Tank Waste Remediation}, \bibinfo{type}{Technical
  Report} \bibinfo{number}{PNNL-8388}, Pacific Northwest National Laboratory,
  \bibinfo{year}{1993}.
\bibitem[{{National Research Council}(1996)}]{NationalResearchCouncil1996}
\bibinfo{editor}{{National Research Council}} (Ed.), \bibinfo{title}{Nuclear
  Wastes: Technologies for Separations and Transmutation},
  \bibinfo{publisher}{The National Academies Press}, \bibinfo{year}{1996}.
  \URLprefix \url{http://www.nap.edu/catalog.php?record_id=4912}.
\bibitem[{DeMuth(1996)}]{DeMuth1996}
\bibinfo{author}{S.~DeMuth}, \bibinfo{title}{Cost Benefit Analysis for Enhanced
  Sludge Washing of Underground Storage Tank High-Level Waste},
  \bibinfo{type}{Technical Report} \bibinfo{number}{LA-UR-96-965}, Los Alamos
  National Laboratory, \bibinfo{year}{1996}.
\bibitem[{{National Research Council}(2001)}]{NationalResearchCouncil2001}
\bibinfo{editor}{{National Research Council}} (Ed.), \bibinfo{title}{Research
  Needs for High-Level Waste Stored in Tanks and Bins at U.S. Department of
  Energy Sites: Environmental Management Science Program},
  \bibinfo{publisher}{The National Academies Press}, \bibinfo{year}{2001}.
  \URLprefix \url{http://www.nap.edu/catalog.php?record_id=10191}.
\bibitem[{Kim et~al.(2011)Kim, Schweiger, Rodriguez, Lepry, Lang, Crum, Vienna,
  Johnson, Marra, and Peeler}]{Kim2011}
\bibinfo{author}{D.-S. Kim}, \bibinfo{author}{M.~J. Schweiger},
  \bibinfo{author}{C.~P. Rodriguez}, \bibinfo{author}{W.~C. Lepry},
  \bibinfo{author}{J.~B. Lang}, \bibinfo{author}{J.~V. Crum},
  \bibinfo{author}{J.~D. Vienna}, \bibinfo{author}{F.~Johnson},
  \bibinfo{author}{J.~C. Marra}, \bibinfo{author}{D.~K. Peeler},
  \bibinfo{title}{Formulation and Characterization of Waste Glasses with
  Varying Processing Temperature}, \bibinfo{type}{Technical Report}
  \bibinfo{number}{PNNL-20774}, Pacific Northwest National Laboratory,
  \bibinfo{year}{2011}. \DOIprefix\doi{10.2172/1028572}.
\bibitem[{Nazarro(2003)}]{Nazarro2003}
\bibinfo{author}{R.~M. Nazarro}, in: \bibinfo{editor}{W.~S. Melfort} (Ed.),
  \bibinfo{booktitle}{Nuclear Waste Disposal: Current Issues and Proposals},
  \bibinfo{publisher}{Nova Science Pub Inc}, \bibinfo{year}{2003}.
\bibitem[{Carreon et~al.(2002)Carreon, Mauss, Johnson, Holton, Wright,
  Peterson, and Rueter}]{Carreon2002}
\bibinfo{author}{R.~Carreon}, \bibinfo{author}{B.~M. Mauss},
  \bibinfo{author}{M.~E. Johnson}, \bibinfo{author}{L.~K. Holton},
  \bibinfo{author}{G.~T. Wright}, \bibinfo{author}{R.~A. Peterson},
  \bibinfo{author}{K.~J. Rueter}, in: \bibinfo{booktitle}{Proceedings of the
  2002 Waste Management Conference}.
\bibitem[{Fiskum et~al.(2009)Fiskum, Draper, MacFarlan, Billing, Edwards,
  Peterson, Buck, Jenson, Shimskey, Daniel, Kozelisky, and Snow}]{Fiskum2009}
\bibinfo{author}{S.~K. Fiskum}, \bibinfo{author}{K.~E. Draper},
  \bibinfo{author}{P.~J. MacFarlan}, \bibinfo{author}{J.~M. Billing},
  \bibinfo{author}{M.~K. Edwards}, \bibinfo{author}{R.~A. Peterson},
  \bibinfo{author}{E.~C. Buck}, \bibinfo{author}{E.~D. Jenson},
  \bibinfo{author}{R.~W. Shimskey}, \bibinfo{author}{R.~C. Daniel},
  \bibinfo{author}{A.~E. Kozelisky}, \bibinfo{author}{L.~A. Snow},
  \bibinfo{title}{Laboratory Demonstration of the Pretreatment Process with
  Caustic and Oxidative Leaching Using Actual Hanford Tank Waste},
  \bibinfo{type}{Technical Report} \bibinfo{number}{PNNL-18007}, Pacific
  Northwest National Laboratory, \bibinfo{year}{2009}.
\bibitem[{Snow et~al.(2007)Snow, Rapko, Poloski, and Peterson}]{Snow2007}
\bibinfo{author}{L.~A. Snow}, \bibinfo{author}{B.~M. Rapko},
  \bibinfo{author}{A.~P. Poloski}, \bibinfo{author}{R.~A. Peterson}, in:
  \bibinfo{booktitle}{Proceedings of the 2007 Waste Management Conference}.
\bibitem[{Vienna et~al.(2014)Vienna, Kim, Schweiger, Piepel, Kroll, and
  Kruger}]{Vienna2014}
\bibinfo{author}{J.~D. Vienna}, \bibinfo{author}{D.~S. Kim},
  \bibinfo{author}{M.~J. Schweiger}, \bibinfo{author}{G.~G. Piepel},
  \bibinfo{author}{J.~O. Kroll}, \bibinfo{author}{A.~A. Kruger},
  \bibinfo{title}{Glass Formulation and Testing for U.S. High-Level Tank
  Wastes}, \bibinfo{type}{Technical Report} \bibinfo{number}{PNNL-SA-84872},
  Pacific Northwest National Laboratory, \bibinfo{year}{2014}.
\bibitem[{Smith et~al.(2011)Smith, Schepens, Blanchard, Shimskey, and
  Peterson}]{Smith2011}
\bibinfo{author}{C.~Smith}, \bibinfo{author}{R.~Schepens},
  \bibinfo{author}{D.~L. Blanchard}, \bibinfo{author}{R.~W. Shimskey},
  \bibinfo{author}{R.~A. Peterson}, in: \bibinfo{booktitle}{Proceedings of the
  2011 Waste Management Conference}.
\bibitem[{Sylvester et~al.(2001)Sylvester, Rutherford, Gonzalez-Martin, Kim,
  Rapko, and Lumetta}]{Sylvester2001}
\bibinfo{author}{P.~Sylvester}, \bibinfo{author}{L.~A. Rutherford},
  \bibinfo{author}{A.~Gonzalez-Martin}, \bibinfo{author}{J.~Kim},
  \bibinfo{author}{B.~M. Rapko}, \bibinfo{author}{G.~Lumetta},
  \bibinfo{journal}{Environ. Sci. Technol.} \bibinfo{volume}{35}
  (\bibinfo{year}{2001}) \bibinfo{pages}{216--221}.
  \DOIprefix\doi{10.1021/es001340n}.
\bibitem[{Lumetta(2008)}]{Lumetta2008}
\bibinfo{author}{G.~J. Lumetta}, \bibinfo{title}{Mechanism of Phosphorus
  Removal from Hanford Tank Sludge by Caustic Leaching},
  \bibinfo{type}{Technical Report} \bibinfo{number}{PNNL-17257}, Pacific
  Northwest National Laboratory, \bibinfo{year}{2008}.
\bibitem[{Kruger et~al.(2010)Kruger, Bowan, Joseph, Gan, Kot, Matlack, and
  Pegg}]{Kruger2010}
\bibinfo{author}{A.~A. Kruger}, \bibinfo{author}{B.~W. Bowan},
  \bibinfo{author}{I.~Joseph}, \bibinfo{author}{H.~Gan}, \bibinfo{author}{W.~K.
  Kot}, \bibinfo{author}{K.~S. Matlack}, \bibinfo{author}{I.~L. Pegg}, in:
  \bibinfo{booktitle}{Proceedings of the 2010 Waste Management Conference}.
\bibitem[{Kruger(2011)}]{Kruger2011}
\bibinfo{author}{A.~A. Kruger}, in: \bibinfo{booktitle}{Proceedings of teh ASME
  2011 14th International Conference on Environmental Remediation and
  Radioactive Waste Management}, p. \bibinfo{pages}{1177}.
  \DOIprefix\doi{10.1115/ICEM2011-59388}.
\bibitem[{Pierce et~al.(2012)Pierce, Hrma, Marcial, Riley, and
  Schweiger}]{Pierce2012}
\bibinfo{author}{D.~A. Pierce}, \bibinfo{author}{P.~Hrma},
  \bibinfo{author}{J.~Marcial}, \bibinfo{author}{B.~J. Riley},
  \bibinfo{author}{M.~J. Schweiger}, \bibinfo{journal}{International Journal of
  Applied Glass Science} \bibinfo{volume}{3} (\bibinfo{year}{2012})
  \bibinfo{pages}{59--68}. \DOIprefix\doi{10.1111/j.2041-1294.2012.00079.x}.
\bibitem[{Smith et~al.(2014)Smith, Lang, Kim, Crum, Schweiger, Crawford, Marra,
  and Vienna}]{Smith2014}
\bibinfo{author}{G.~L. Smith}, \bibinfo{author}{J.~B. Lang},
  \bibinfo{author}{D.~Kim}, \bibinfo{author}{J.~V. Crum},
  \bibinfo{author}{M.~J. Schweiger}, \bibinfo{author}{C.~L. Crawford},
  \bibinfo{author}{J.~C. Marra}, \bibinfo{author}{J.~D. Vienna},
  \bibinfo{title}{Silicate Based Glass Formulations for Immobilization of U.S.
  Defense Wastes Using Cold Crucible Induction Melters},
  \bibinfo{type}{Technical Report} \bibinfo{number}{PNNL-23288}, Pacific
  Northwest National Laboratory, \bibinfo{year}{2014}.
\bibitem[{Aloise(2009)}]{Aloise2009}
\bibinfo{author}{G.~Aloise}, \bibinfo{title}{Nuclear Waste: Uncertainties and
  Questions about Costs and Risks Persist with DOE?s Tank Waste Cleanup
  Strategy at Hanford}, \bibinfo{type}{Technical Report}
  \bibinfo{number}{GAO-09-913}, United States Governement Accountability
  Office, \bibinfo{year}{2009}.
\bibitem[{Lehnert(1971)}]{Lehnert1971}
\bibinfo{author}{B.~Lehnert}, \bibinfo{journal}{Nuclear Fusion}
  \bibinfo{volume}{11} (\bibinfo{year}{1971}) \bibinfo{pages}{485--}.
  \DOIprefix\doi{10.1088/0029-5515/11/5/010}.
\bibitem[{Krishnan et~al.(1981)Krishnan, Geva, and Hirshfield}]{Krishnan1981}
\bibinfo{author}{M.~Krishnan}, \bibinfo{author}{M.~Geva},
  \bibinfo{author}{J.~L. Hirshfield}, \bibinfo{journal}{Phys. Rev. Lett.}
  \bibinfo{volume}{46} (\bibinfo{year}{1981}) \bibinfo{pages}{36--38}.
  \DOIprefix\doi{10.1103/PhysRevLett.46.36}.
\bibitem[{Grossman and Shepp(1991)}]{Grossman1991}
\bibinfo{author}{M.~W. Grossman}, \bibinfo{author}{T.~A. Shepp},
  \bibinfo{journal}{IEEE Transactions on Plasma Science} \bibinfo{volume}{19}
  (\bibinfo{year}{1991}) \bibinfo{pages}{1114--1122}.
  \DOIprefix\doi{10.1109/27.125034}.
\bibitem[{Rax et~al.(2007)Rax, Robiche, and Fisch}]{Rax2007}
\bibinfo{author}{J.-M. Rax}, \bibinfo{author}{J.~Robiche},
  \bibinfo{author}{N.~J. Fisch}, \bibinfo{journal}{Phys. Plasmas}
  \bibinfo{volume}{14} (\bibinfo{year}{2007}) \bibinfo{pages}{043102--8}.
  \DOIprefix\doi{10.1063/1.2717882}.
\bibitem[{Freeman et~al.(2003)Freeman, Agnew, Anderegg, Cluggish, Gilleland,
  Isler, Litvak, Miller, O'Neill, Ohkawa, Pronko, Putvinski, Sevier, Sibley,
  Umstadter, Wade, and Winslow}]{Freeman2003}
\bibinfo{author}{R.~Freeman}, \bibinfo{author}{S.~Agnew},
  \bibinfo{author}{F.~Anderegg}, \bibinfo{author}{B.~Cluggish},
  \bibinfo{author}{J.~Gilleland}, \bibinfo{author}{R.~Isler},
  \bibinfo{author}{A.~Litvak}, \bibinfo{author}{R.~Miller},
  \bibinfo{author}{R.~O'Neill}, \bibinfo{author}{T.~Ohkawa},
  \bibinfo{author}{S.~Pronko}, \bibinfo{author}{S.~Putvinski},
  \bibinfo{author}{L.~Sevier}, \bibinfo{author}{A.~Sibley},
  \bibinfo{author}{K.~Umstadter}, \bibinfo{author}{T.~Wade},
  \bibinfo{author}{D.~Winslow}, \bibinfo{journal}{AIP Conf. Proc.}
  \bibinfo{volume}{694} (\bibinfo{year}{2003}) \bibinfo{pages}{403--410}.
  \DOIprefix\doi{10.1063/1.1638067}.
\bibitem[{Gueroult and Fisch(2014)}]{Gueroult2014a}
\bibinfo{author}{R.~Gueroult}, \bibinfo{author}{N.~J. Fisch},
  \bibinfo{journal}{Plasma Sources Science and Technology} \bibinfo{volume}{23}
  (\bibinfo{year}{2014}) \bibinfo{pages}{035002--}.
  \DOIprefix\doi{10.1088/0963-0252/23/3/035002}.
\bibitem[{Ohkawa and Miller(2002)}]{Ohkawa2002}
\bibinfo{author}{T.~Ohkawa}, \bibinfo{author}{R.~L. Miller},
  \bibinfo{journal}{Phys. Plasmas} \bibinfo{volume}{9} (\bibinfo{year}{2002})
  \bibinfo{pages}{5116--5120}. \DOIprefix\doi{10.1063/1.1523930}.
\bibitem[{Fetterman and Fisch(2011)}]{Fetterman2011}
\bibinfo{author}{A.~J. Fetterman}, \bibinfo{author}{N.~J. Fisch},
  \bibinfo{journal}{Phys. Plasmas} \bibinfo{volume}{18} (\bibinfo{year}{2011})
  \bibinfo{pages}{094503--3}. \DOIprefix\doi{10.1063/1.3631793}.
\bibitem[{Gueroult and Fisch(2012)}]{Gueroult2012a}
\bibinfo{author}{R.~Gueroult}, \bibinfo{author}{N.~J. Fisch},
  \bibinfo{journal}{Phys. Plasmas} \bibinfo{volume}{19} (\bibinfo{year}{2012})
  \bibinfo{pages}{122503--6}. \DOIprefix\doi{10.1063/1.4771674}.
\bibitem[{Gueroult et~al.(2014)Gueroult, Rax, and Fisch}]{Gueroult2014}
\bibinfo{author}{R.~Gueroult}, \bibinfo{author}{J.-M. Rax},
  \bibinfo{author}{N.~J. Fisch}, \bibinfo{journal}{Physics of Plasmas}
  \bibinfo{volume}{21} (\bibinfo{year}{2014}) \bibinfo{pages}{020701}.
  \DOIprefix\doi{10.1063/1.4864325}.
\bibitem[{Fetterman and Fisch(2011)}]{Fetterman2011b}
\bibinfo{author}{A.~J. Fetterman}, \bibinfo{author}{N.~J. Fisch},
  \bibinfo{journal}{Phys. Plasmas} \bibinfo{volume}{18} (\bibinfo{year}{2011})
  \bibinfo{pages}{103503--8}. \DOIprefix\doi{10.1063/1.3646311}.
\bibitem[{Tanaka et~al.(2007)Tanaka, Pigarov, Smirnov, Krasheninnikov, Ohno,
  and Uesugi}]{Tanaka2007}
\bibinfo{author}{Y.~Tanaka}, \bibinfo{author}{A.~Y. Pigarov},
  \bibinfo{author}{R.~D. Smirnov}, \bibinfo{author}{S.~I. Krasheninnikov},
  \bibinfo{author}{N.~Ohno}, \bibinfo{author}{Y.~Uesugi},
  \bibinfo{journal}{Physics of Plasmas} \bibinfo{volume}{14}
  (\bibinfo{year}{2007}) \bibinfo{pages}{052504}.
  \DOIprefix\doi{10.1063/1.2722274}.
\bibitem[{McGinnis et~al.(1999)McGinnis, Welch, and Hunt}]{McGinnis1999}
\bibinfo{author}{C.~P. McGinnis}, \bibinfo{author}{T.~D. Welch},
  \bibinfo{author}{R.~D. Hunt}, \bibinfo{journal}{Separation Science and
  Technology} \bibinfo{volume}{34} (\bibinfo{year}{1999})
  \bibinfo{pages}{1479--1494}. \DOIprefix\doi{10.1080/01496399908951104}.
\bibitem[{Perez et~al.(2001)Perez, Peeler, Bickford, Strachan, Day, Triplett,
  Kim, Vienna, Lambert, Wittman, and Marra}]{Perez2001}
\bibinfo{author}{J.~M. Perez}, \bibinfo{author}{D.~K. Peeler},
  \bibinfo{author}{D.~F. Bickford}, \bibinfo{author}{D.~M. Strachan},
  \bibinfo{author}{D.~E. Day}, \bibinfo{author}{M.~B. Triplett},
  \bibinfo{author}{D.~S. Kim}, \bibinfo{author}{J.~D. Vienna},
  \bibinfo{author}{S.~I. Lambert}, \bibinfo{author}{R.~S. Wittman},
  \bibinfo{author}{S.~L. Marra}, \bibinfo{title}{High Level Waste Melter Study
  Report}, \bibinfo{type}{Technical Report} \bibinfo{number}{PNNL-13582},
  Pacific Northwest National Laboratory, \bibinfo{year}{2001}.
\bibitem[{Chew and Hamm(2014)}]{Chew2014}
\bibinfo{author}{D.~P. Chew}, \bibinfo{author}{B.~A. Hamm},
  \bibinfo{title}{Liquid Waste System Plan Revision 19},
  \bibinfo{type}{Technical Report} \bibinfo{number}{SRR-LWP-2009-00001},
  Savannah River Remediation LLC, \bibinfo{year}{2014}.
\bibitem[{Mazhukin et~al.(2007)Mazhukin, Nossov, and Smurov}]{Mazhukin2007}
\bibinfo{author}{V.~I. Mazhukin}, \bibinfo{author}{V.~V. Nossov},
  \bibinfo{author}{I.~Smurov}, \bibinfo{journal}{Journal of Applied Physics}
  \bibinfo{volume}{101} (\bibinfo{year}{2007}) \bibinfo{pages}{024922}.
  \DOIprefix\doi{10.1063/1.2431951}.
\bibitem[{Samsonov(1973)}]{Samsonov1973}
\bibinfo{author}{G.~Samsonov}, in: \bibinfo{editor}{G.~Samsonov} (Ed.),
  \bibinfo{booktitle}{The Oxide Handbook}, \bibinfo{publisher}{Springer US},
  \bibinfo{year}{1973}, pp. \bibinfo{pages}{36--223}.
  \DOIprefix\doi{10.1007/978-1-4615-9597-7_3}.
\bibitem[{Lieberman and Lichtenberg(1994)}]{Lieberman1994}
\bibinfo{author}{M.~A. Lieberman}, \bibinfo{author}{A.~J. Lichtenberg},
  \bibinfo{title}{Principles of Plasma Discharge for Materials Processing},
  \bibinfo{publisher}{John Wiley \& Sons}, \bibinfo{year}{1994}.
\bibitem[{Harrington(2011)}]{Harrington2011}
\bibinfo{author}{S.~J. Harrington}, \bibinfo{title}{Compilation of Laboratory
  Scale Aluminum Wash and Leach Report Results}, \bibinfo{type}{Technical
  Report} \bibinfo{number}{RPP-RPT-46791}, Washington River Protection
  Solutions, LLC, \bibinfo{year}{2011}. \DOIprefix\doi{10.2172/1004086}.
\bibitem[{Rapko and Vienna(2002)}]{Rapko2002}
\bibinfo{author}{B.~M. Rapko}, \bibinfo{author}{J.~D. Vienna},
  \bibinfo{title}{Selective Leaching of Chromium from Hanford Tank Sludge
  241-U-108}, \bibinfo{type}{Technical Report} \bibinfo{number}{PNNL-14019},
  Pacific Northwest National Laboratory, \bibinfo{year}{2002}.
  \DOIprefix\doi{10.2172/860129}.
\bibitem[{Geeting and Hallen(2005)}]{Geeting2005}
\bibinfo{author}{J.~G.~H. Geeting}, \bibinfo{author}{R.~T. Hallen},
  \bibinfo{journal}{Separation Science and Technology} \bibinfo{volume}{40}
  (\bibinfo{year}{2005}) \bibinfo{pages}{1--15}.
  \DOIprefix\doi{10.1081/SS-200041752}.

\end{thebibliography}

\end{document}